\documentclass[aps,pre,showpacs,twocolumn,superscriptaddress,floatfix]{revtex4}
\usepackage{graphicx}
\usepackage{epsfig} 
\usepackage{epstopdf}
\usepackage{color}
\usepackage{amsfonts}
\usepackage{amssymb,amsmath}
\usepackage{bbold}

\begin{document}
 
\title{From synchronous to one-time delayed dynamics in coupled maps}
 
\author{Celia Anteneodo}
\email{celia.fis@puc-rio.br}
\author{Juan Carlos Gonz\'alez-Avella}
\email{avellaj@gmail.com}
\affiliation{Department of Physics, PUC-Rio, Caixa Postal 38071,
             22452-970 Rio de Janeiro, Brazil} 
\author{Ra\'ul O. Vallejos}
\email{vallejos@cbpf.br}
\affiliation{Centro Brasileiro de Pesquisas F\'{\i}sicas (CBPF), Rua Dr.~Xavier Sigaud 150, 
             22290-180 Rio de Janeiro, Brazil}
\date{\today}

\begin{abstract}
We study the completely synchronized states (CSSs) of a system of coupled logistic maps 
as a function of three parameters:
interaction strength ($\varepsilon$), 
range of the interaction ($\alpha$), that can vary from first-neighbors to global coupling, 
and a parameter ($\beta$) that allows to scan continuously from non-delayed to one-time delayed dynamics.
We identify in the plane $\alpha$-$\varepsilon$ periodic orbits, limit cycles and chaotic trajectories, 
and describe how these structures change with the delay.
These features can be explained by studying the bifurcation diagrams of a two-dimensional non-delayed map. 
This allows us to understand the effects of one-time delays on CSSs,  
e.g, regularization of chaotic orbits and synchronization of short-range coupled maps, 
observed when the dynamics is moderately delayed.
Finally, we substitute the logistic map by cubic and logarithmic maps, in order to test the
robustness of our findings.
\end{abstract}

\maketitle

\section{Introduction}

Coupled map lattices are paradigmatic models for the study of complex collective 
behavior such as synchronization~\cite{PykovskyBook,KanekoBook,ManrubiaBook,KuramotoBook}. 
This collective phenomenon presents scientific and technological interest for as diverse areas
as Josephson junction arrays, multimode lasers, vortex dynamics, even evolutionary biology~\cite{PykovskyBook}, 
and also constitute good prototypes to investigate control of chaos. 

The asymptotic collective states can be highly influenced by asynchronicities, which are 
present in real systems~\cite{Sinha2007,Sinha_2000,Lumer1994,nuno1,nuno2}.  
For instance, asynchronous updating may open windows in parameter space where synchronization 
becomes allowed~\cite{Sinha_2000} and induce regularity in coupled 
systems in contrast to a synchronous updating~\cite{Sinha2007,unstableFP1,squeeze}.  
Similarly, the introduction of time delays, to account for finite propagation times 
in information transmission among units~\cite{Marti_PRL05,Marti_EPJB05,Marti_PRE05,delayed},    
has also a noticeable impact on the collective patterns, e.g.,  
although synchronization is still possible, chaos may be suppressed~\cite{Marti_PRL05}. 

Within the spatial domain, a realistic ingredient that has important impact on the 
determination of the emergent patterns in extended systems is the coupling 
range~\cite{chemical1,Anteneodo2004,Anteneodo2006,Marti_EPJB05,gallas2004,celia1998,celia2000,lind04,Avella_PRE16}.  
Insofar as the range of the interactions can affect the propagation 
of information, it is important to explore its interplay with the updating scheme. 

In this work we propose a scheme of coupled maps that integrates these two characteristics, 
allowing  to explore its interplay. 
Namely, the coupling depends both on the distance among units, covering from nearest-neighbors to 
global interaction, as well as on a delay protocol 
that scans continuously from the fully synchronous to the fully one-time delayed dynamics.
In particular, we monitor the emergence and breakdown of phase space structures, 
such as fixed points, periodic orbits, limit cycles, etc., in completely synchronized states (CSSs),  
when the contribution of delays and the range of the interactions 
change  between  extreme cases.


\section{The Model} 
\label{sec:model}

We consider a linear chain of $N$ maps with periodic boundary conditions.  
The maps are coupled according to the diffusive scheme and obey the following
dynamical equations
 \begin{equation}
   x_{t+1}^i = (1-\varepsilon) f(x_t^i) +  \varepsilon 
   \sum_{r=1}^{N'} A(r) \Bigl( f(\hat{x}_{t}^{i-r}) +
     f(\hat{x}_{t}^{i+r})  \Bigr) \, ,
   \label{CMLbeta}
 \end{equation}
for  $i=1,\ldots,N$. 
Here $x_t^i$ describes the state of map $i$ at discrete time $t$, whose uncoupled 
dynamics is governed by the chaotic map $f(x)$. 
Delayed coordinates $\hat{x}_t^j$ are defined by
\begin{equation}
\hat{x}_t^j= \beta x_{t-1}^j + (1-\beta) x_{t}^j \, .
\label{beta}
\end{equation}
Thus, the interpolation parameter $\beta$ ($0 \le \beta \le 1$)  
allows to scan continuously from the synchronous or non-delayed dynamics ($\beta=0$), 
to the pure one-time delayed dynamics ($\beta=1$). 
Equation~(\ref{CMLbeta}) describes a fully connected array where elements interact 
through a coupling which depends on $f(x)$, 
with intensity $A(r)$, where $r$ is the integer distance between maps on the ring. 
Finally, the coupling parameter $\varepsilon$ ($0\le \varepsilon \le 1$) sets the relative weights 
of global and local influences. 
At each time step, the $N$ maps are updated simultaneously.

In numerical examples, we will consider $A(r)= r^{-\alpha}/\eta$,   
where $\alpha \in [0,\infty)$ determines the range of the interactions, 
and 
$\eta(\alpha) = 2\sum_{r=1}^{N'}r^{-\alpha}$ 
is a normalization factor, with 
$N' = (N-1)/2$ for odd $N$.
This coupling scheme allows to switch smoothly from  global  ($\alpha = 0$) 
to  nearest-neighbor ($\alpha \rightarrow \infty$) interactions. 

In numerical simulations we will use, as paradigmatic example,  
mainly the  logistic map	$f(x)=4x(1-x)$. 
However, we will also show results for logarithmic and cubic maps. 
In analytical expressions we will keep the generic forms of 
$A(r)$ and $f(x)$ as far as possible.

\section{Results} 
\label{sec:results}

We performed numerical simulations of the coupled maps lattice defined by Eq.~(\ref{CMLbeta}),  
with the forms of $A(r)$ and $f(x)$ described above,
starting from random initial conditions (uniform in the inteval [0,1]). 
We considered different values for the coupling strength $\varepsilon$, 
the range of the interactions ruled by $\alpha$, and 
the time delay parameter $\beta$.
 
We monitored the collective behavior by means of the
instantaneous mean field $h_t$ defined as~\cite{Marti_PRL05,lind04} 
\begin{equation}
    h_t = \frac{1}{N}\sum_{i=1}^N  x^i_t  \,.
\end{equation}
In order to measure the degree of  synchronization,    
we use the  time average, $ \langle \sigma_t \rangle$, 
of the instantaneous standard deviation of $h_t$, namely,   
\begin{equation}
      \sigma_t = \sqrt{ \frac{1}{N}\sum_{i=1}^N 
			\bigl(    x^i_t   -h_t \bigr)^2 } \,. 
 \end{equation}
When  
$\sigma \equiv \langle \sigma_t \rangle = 0 $, 
it means that the system is completely synchronized, i.e, 
for all $t$, it holds  
$x^1_t=x^2_t=\ldots x^N_t=x^\star_t$, 
where $x^\star_t$ can trace a chaotic or a regular trajectory. 
In the latter case, we measured the period of the short orbits. 

Another relevant parameter, which allows to characterize the orbits of a 
dynamical system, is the largest Lyapunov exponent $\lambda_{\rm max}$~\cite{eckman-ruelle}. 
If $\lambda_{\rm max}$ is  positive, the system displays a chaotic behavior, 
while if it is  negative, the dynamics is regular.  
We computed $\lambda_{\rm max}$ using the Benettin algorithm~\cite{Benettin_01}.  
 
\begin{center}
\begin{figure}[t!]
\includegraphics[width=1.0\linewidth,angle=0]{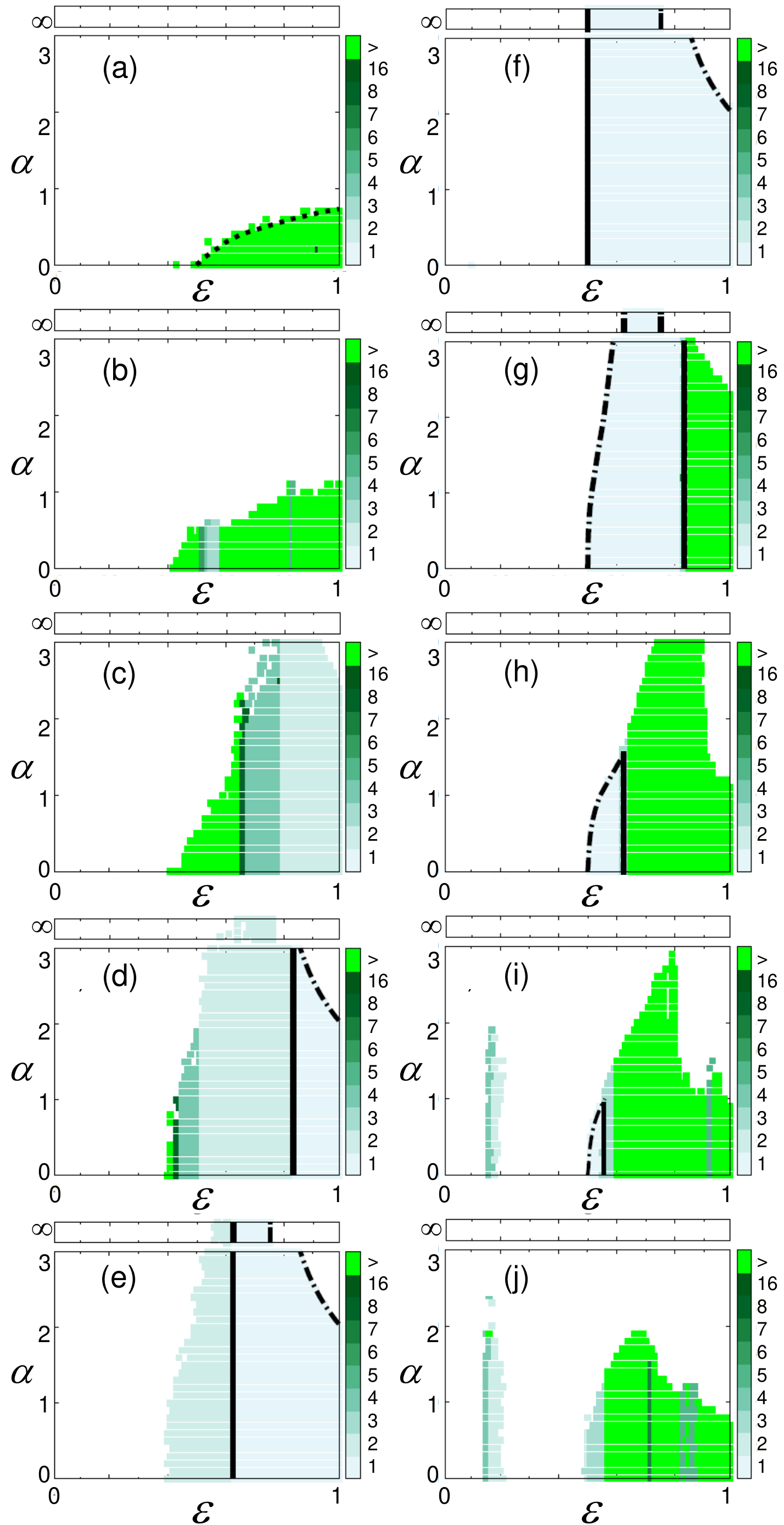}
\caption{ (Color online)  
Phase diagrams in the plane $\varepsilon-\alpha$ for $\beta=$ 
0.0 (a), 0.1 (b), 0.2 (c), 0.3 (d), 0.4 (e), 0.5 (f), 0.6 (g), 0.8 (h), 0.9 (i) 
and 1.0 (j). 
Only regions where coupled maps fully synchronize are colored. 
The color scale indicates the period of the synchronized orbits.  
The local dynamics corresponds to the logistic map $f(x)=4x(1-x)$.
The black lines indicate the frontiers of the regions of period-1 orbits: 
vertical solid lines are given by the longitudinal stability condition (\ref{cond1}) 
and point-dashed lines by the transverse stability condition (\ref{eq:stab1}). 
The dotted line in the first panel is given by Eq.~(\ref{frontiers}).
In all cases, the array size is $N=201$.   
We chose initial conditions, $x^i_0$ and $x^i_1$, randomly in [0,1]. 
 }
\label{fig:PD1}
\end{figure}
\end{center}

In Fig.~\ref{fig:PD1}, we present  phase diagrams in the plane $\varepsilon-\alpha$ 
obtained from numerical simulations, for the local dynamics given by the logistic 
map $f(x)=4x(1-x)$, and for different values of $\beta$. 
For each pair $(\varepsilon,\alpha)$ we calculated $\sigma$ by averaging over 100 time steps, 
after a transient of $t>10^3$. 
We considered the state as fully synchronized if $\sigma < 10^{-3}$.
Only the parameter regions where complete synchronization occurs are colored. 
Shown are regions containing periodic orbits up to period-16. 
Green regions may contain periodic (period $>$ 16), quasi-periodic, intermittent or chaotic
trajectories (see below).  
%
%
%
%
Figure~\ref{fig:PD1} shows several known features such as the existence of an 
interval of the coupling strength $\varepsilon$ for the system to synchronize, 
and that long-range interactions favor synchronization, i.e.,   
stability of completely synchronized (CS) regions shrink as $\alpha$ grows.  
Besides these features, Fig.~\ref{fig:PD1} also exhibits important changes with $\beta$. 
In particular, there is a qualitative change of the stability regions around $\beta \approx 0.5$.

Let us first look the case $\beta=0.0$ in Fig.~\ref{fig:PD1}(a): 
here CSSs are fully chaotic; however these states only exist
above minimum values of 1) interaction strength and 2) interaction range.  
By increasing the contribution of delayed states ($\beta>0$), 
windows of regular behavior are introduced. 
For instance,  in the case $\beta=0.1$ the visible windows of regularity correspond to 
period doubling cascades (from right to left) of periods $3\times 2^k$ 
and $5\times 2^k$, with $k=0,1,2,\ldots$, 
around $\epsilon \simeq 0.55$ and 0.83, respectively. 
In the case $\beta=0.2$, shown in panel (c), we neatly notice  a period doubling cascade from right to left
of periods $2^k$, with $k=1,2,\ldots$ leading to chaos below $\varepsilon \approx 0.67$. 
This cascade is shifted to the left as $\beta$ increases.
The larger the delay contribution, the shorter range of the interactions that gives rise to CSSs.
Moreover, when $\beta \in  (\approx 0.3,\approx 0.6)$, i.e., moderate delay contribution, 
even nearest-neighbor couplings allow complete synchronization of regular trajectories. 
When the contribution of delays is large enough, the structure of bifurcations becomes more complex, 
and will be discussed in Sec.~\ref{sec:2Dmap}.

These qualitative changes,   that for the logistic map occur around $\beta \approx 0.5$, 
can be understood heuristically as follows. 
When complete synchronization occurs, all maps share the same state $x^\star_t$ at each time step, 
then,  substituting $x^i_t$ by  $x^\star_t$ into Eq.~(\ref{CMLbeta}), the summation becomes   
$\varepsilon f(\hat{x}^\star_t)$. Thus, the dependence on $\alpha$ and $N$ embodied in $\eta$ cancels out. 
Finally, using the definition of $\hat{x}^\star_t$ given by Eq.~(\ref{beta}), 
Eq.~(\ref{CMLbeta}) becomes
 \begin{equation}
   x^\star_{t+1} = (1-\varepsilon) f(x^\star_t) +  \varepsilon f(\beta x^\star_{t-1}+\{1-\beta\} x^\star_t) .
   \label{2Dmap}
 \end{equation}
The stability of a given trajectory $x^\star_t$ depends only on $\varepsilon$ and $\beta$.
However, when thought of as a CSS of the coupled-map lattice, 
its stability is also governed by $\alpha$ and $N$, while 
parameters $\varepsilon$ and $\beta$ only determine the longitudinal stability of CSSs.

For weakly delayed dynamics, the map (\ref{2Dmap}) resembles the logistic map and its bifurcation diagram is logistic-like.
However, for delay dominated dynamics ($\beta > 0.5$ in this case), 
the dynamics of CSSs becomes really two-dimensional 
and the family of bifurcations expands. 
We return to this issue in Sec.~\ref{sec:2Dmap}.

Now, let us analyze the stability of CSSs, starting by periodic motion.

\subsection{Stability of the CSS: fixed point}
\label{sec:fixedpoint}

If the system synchronizes at a fixed point $X^0$, then for each element $i$, 
one has 
$x^i_t=x^i_{t-1}=x^\star_t=X^0$, 
hence 
$\hat{x}_t^i= \beta x_{t-1}^i + (1-\beta) x_{t-0}^i=X^0$. 
Since, from Eq.~(\ref{CMLbeta}), 
each equation in the array verifies the single map relation  $ X^0 = f(X^0)$, 
then the fixed points of the array are identical to the fixed points of the uncoupled maps.
For instance, in the case $f(x)=4x(1-x)$, 
the fixed points are $X^0=0$ and $X^0=3/4$, which are unstable in the uncoupled map, 
since in both cases $|f^\prime(X^0)|>1$. 
However, their stability is expected to change in the array. 

In order to study the stability (both longitudinal and transversal) of the fixed points of the coupled system, 
we consider the linearized form of Eq.~(\ref{CMLbeta}) around a fixed point $X^0$, which reads  
 \begin{equation}
   \delta x_{t+1}^i =  (1-\varepsilon) D^0 \delta x_{t}^i +  \varepsilon   \sum_{r=1}^{N'} A(r) D^0  
	\Bigl( \delta \hat{x}_t^{i-r}  +   \delta \hat{x}_{t}^{i+r} \Bigr),
   \label{lin}
 \end{equation}
where $D^0\equiv f^\prime(X^0)$ and $\delta\hat{x}_t^j= \beta \delta {x}_{t-1}^j + (1-\beta) \delta {x}_{t-0}^j$.
%
The system of linear equations (\ref{lin}) can be cast in the  form  
 \begin{equation}
 \mathbf{\delta \tilde{x}}_{t+1} = \mathbf{F^0 }\mathbf{\delta \tilde{x}}_{t}    
\label{CMLmatrix}
 \end{equation}
by defining the $2N$ dimensional tangent vector 
$\mathbf{\delta \tilde{x}}_t=(\mathbf{ \delta x}_t,\mathbf{\delta x}_{t-1})^t$, 
and  the $2N \times 2N$ (time independent) matrix $\mathbf{F^0 }$:  
\begin{equation}
 \mathbf{ F^0}  = \left( 
\begin{array}{c|c}
 D^0[(1-\varepsilon)\mathbb{1}+(1-\beta)\varepsilon \mathbf{A}]    & D^0\beta\varepsilon \mathbf{A}  \\
\hline
                \mathbb{1}                          &\mathbb{0}  
\end{array} 
\right) ,
\label{tildeF}
\end{equation}
where $\mathbf{A}$ is the $N\times N$ matrix  whose elements are  $A_{ij}=(1-\delta_{ij}) A( r_{ij})$
and $r_{ij}$ is the distance between elements $i$ and $j$ on the circle, i.e., 
$r_{ij}={\rm min}_k|i-j+kN|$. 
The eigenvalues of $\mathbf A$ are obtained by Fourier diagonalization~\cite{Anteneodo2004}: 
\begin{equation}
a_k=2\sum_{m=1}^{N'} A(m) \cos(2\pi km/N) \, , 
\end{equation}
for $1\le k \le N$.  
By setting $ \mathbf{ F^0}  ( \mathbf{u}, \mathbf{v})^t = \lambda ( \mathbf{u}, \mathbf{v})^t $, 
and using Eq.~(\ref{tildeF}), we obtain 
\begin{eqnarray} \nonumber
D^0[(1-\varepsilon)\mathbb{1}+(1-\beta)\varepsilon \mathbf{A}]\mathbf{u} + 
D^0\beta\varepsilon \mathbf{A}\mathbf{v}&=&\lambda \mathbf{u}\\
\mathbf{u} &=& \lambda \mathbf{v} \,.
\end{eqnarray}
Simple algebra shows that the eigenvalues $\lambda$ of the block matrix $\mathbf{F^0}$ 
are related to the eigenvalues $a_k$ ($k=1,\ldots,N$) of $\mathbf{A}$, 
through the characteristic equations:
\begin{equation}
\lambda^2 -\lambda D^0[  1-\varepsilon +\varepsilon(1-\beta)a_k] -D^0\beta\varepsilon a_k = 0 \, ,
\label{characteristic}
\end{equation}
each $a_k$ giving two values of $\lambda$.
The conditions $|\lambda|< 1$ for all eigenvalue $a_k$ determine the region of stability of the fixed point
in the parameter space $(\alpha,\beta,\varepsilon)$. 
The spectrum of $\mathbf{A}$ has been analyzed elsewhere \cite{Anteneodo2004}. 
The largest eigenvalue of $\mathbf{A}$ is $a_N=1$ for all $\alpha$. 
The corresponding eigenvector is proportional to $\mathbf{e}_N=(1,1,1,\ldots,1)$, that is, 
it belongs to the completely synchronized subspace. 
Therefore, $|\lambda_N|<1$ is the condition for longitudinal stability (along the CS subspace). 
The remaining eigenvalues provide the conditions for transverse stability.  

For the logistic map, the fixed point $X^0=0$ remains longitudinally unstable in the array 
for any $(\varepsilon,\beta)$. For   $X^0=3/4$, using $a_N=1$ in Eq.~(\ref{characteristic}), 
and setting $|\lambda|<1$, we obtain the condition 
\begin{equation} \label{cond1}
1/4 < \beta \varepsilon < 1/2,
\end{equation} 
which guarantees longitudinal stability. Therefore, depending on $\varepsilon$ and $\beta$, this fixed point 
can gain stability. 
In particular for $\beta=0$ (no delay), the fixed point $X^0=3/4$ cannot be stable 
for any coupling strength $\varepsilon$, 
but as $\beta$ grows beyond $\beta = 1/4$, there appears an interval of 
values of $\varepsilon$ for which the fixed point becomes stable.  
This shows the emergence of the longitudinal stability 
of the locally unstable fixed point as the delay increases.
Concerning transverse stability, 
we checked numerically that eigenvalues $0<a_k<1$ do not add restrictions 
to the region of stability defined by $a_N=1$. 
Eigenvalues $a_{\rm min}<a_k<0$ set the frontier:
\begin{equation}
  \frac{1}{2 [ 1 - a_{\rm min} (1 -2 \beta  ) ]} <  \varepsilon <  \frac{3}{2 ( 1 - a_{\rm min}  )} \, ,  
\label{eq:stab1}
\end{equation}
recalling that $a_{\rm min}=a_{N/2}$ (for $N$ odd).
This restriction depends both on $\alpha$ and $N$ through $a_{\rm min}$. 
However, the $N$-dependence is very weak for $N$ large, e.g., $N$=100.
These frontiers are represented by lines in Fig.~\ref{fig:PD1}. 

Equation~(\ref{cond1}) shows how the contribution of delays, through $\beta$, directly influences the 
longitudinal stability of fixed points, which occurs for a moderate range of values of $\beta$. 
Moreover, Eq.~(\ref{eq:stab1}) shows how $\beta$ influences the lower bound of transverse stability, 
turning complete synchronization at the fixed point visible. 
This condition is relevant only for $\beta>0.5$, because the condition of longitudinal stability 
$\varepsilon > 1/(4\beta)$ is more restrictive in such case.

Complete synchronization at the fixed point $X^0=3/4$ can be seen in Fig.~\ref{fig:ht}(a), 
for  $\beta=0.3$, and in Fig.~\ref{fig:ht}(b) for $\beta=0.8$, when $\alpha=0.5$.
The intervals of stability as a function of $\varepsilon$ coincide with
the analytical calculation, i.e., 
$\varepsilon>0.833$ ($\beta=0.3$) and $0.510<\varepsilon<0.625$ ($\beta=0.8$).

\subsection{Stability of CS period-2 trajectories}

A period-2 trajectory of the coupled system (with values $X^1$ and $X^2$) must verify
\begin{eqnarray} \nonumber
X^1 &=& (1-\varepsilon) f(X^2) + \varepsilon f( \beta X^1 +[1-\beta]X^2) \, ,\\
X^2 &=& (1-\varepsilon) f(X^1) + \varepsilon f( \beta X^2 +[1-\beta]X^1) \, .
\end{eqnarray}
For the logistic map, besides the trivial solutions $(X^1,X^2)=(0,0)$ and $(3/4,3/4)$, 
there are the (equivalent) nontrivial solutions 
$(X^1,X^2)$, $(X^2,X^1)$ with $X^1\neq X^2$, which are a function of $\varepsilon$ and $\beta$. 
In order to study the stability of period-2 orbits, we write the evolution of the tangent vectors as
\begin{equation}
\mathbf{\delta \tilde{x}}_{t+2} = \mathbf{F^1 }\mathbf{F^2 }\mathbf{\delta \tilde{x}}_{t}    
\label{CMLmatrix}
\end{equation}
where
\begin{equation}
\mathbf{ F^k}  = \left( 
\begin{array}{c|c}
 D^k(1-\varepsilon)\mathbb{1}+\tilde{D}^k(1-\beta)\varepsilon \mathbf{A}]  & \tilde{D}^k\beta\varepsilon \mathbf{A}  \\
\hline
                 \mathbb{1}                                &\mathbb{0}  
\end{array} 
\right) \, ,
\label{tildeF}
\end{equation}
with $D^k=f^\prime(X^k)$ and $\tilde{D}^k=f^\prime(\beta X^j +[1-\beta]X^k)$,  for $k=1,2$ and $j\neq k$.
Again, the condition $|\lambda|< 1$ for all $a_k$ (with $\lambda$ eigenvalue of $\mathbf{F^2 }\mathbf{F^1 }$) 
provides the region of stability of the period-2 orbits. 
This case, however, is not amenable to analytical treatment because the trajectories must be found
numerically.

\begin{center}
\begin{figure}[b!]
\includegraphics[width=0.8\linewidth,angle=0]{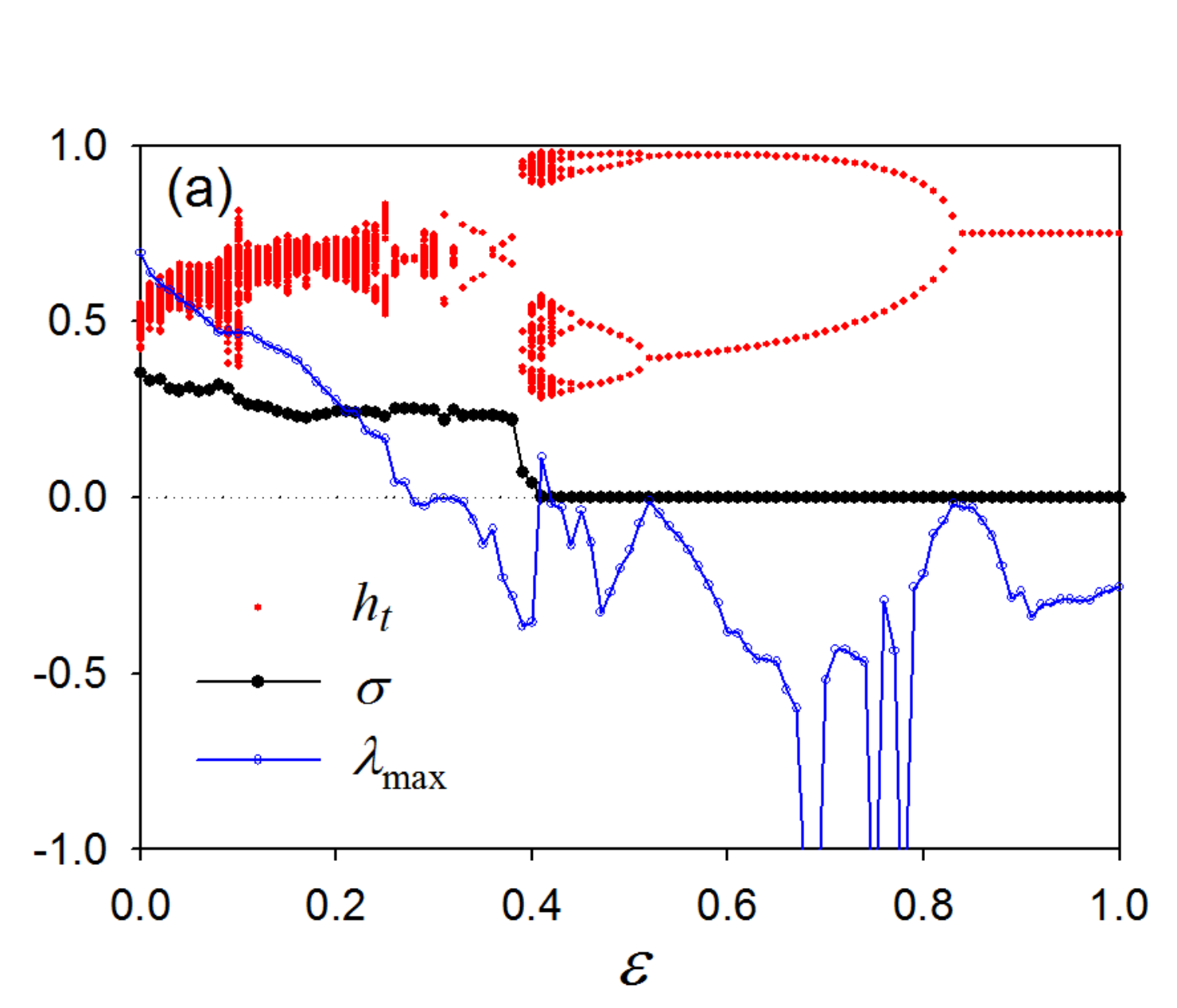}
\includegraphics[width=0.8\linewidth,angle=0]{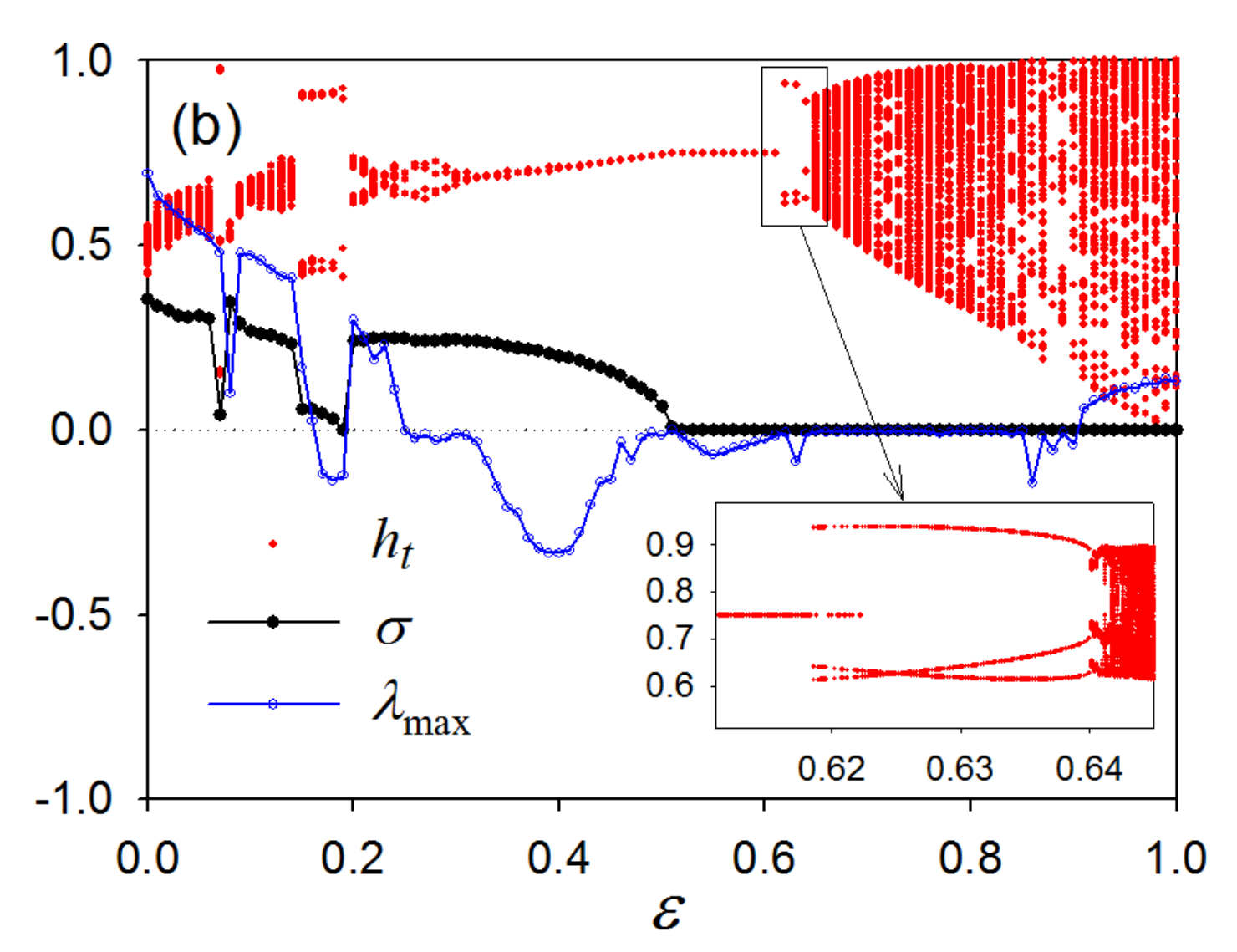}
\caption{ (Color online)  
Order parameter $\sigma$ (black dots), mean field $h_t$ 
(red light dots; we plot 100 consecutive values after a transient of $t>10^3$ ) and 
largest Lyapunov exponent $\lambda_{max}$ vs. coupling $\varepsilon$, 
for $\alpha=0.5$, with $\beta=0.3$  (a), $\beta=0.8$  (b). 
The local dynamics is given by the logistic map. 
In all cases the array size is  $N=201$. 
The inset is a zoom of the main frame to show the coexistence of period-1 and period-3 orbits. 
At the point where two branches of period-3 orbits cross, two of the three 
values  become equal, but remaining period-3. 
This is possible because the state of the system depends on two previous times.
} 
\label{fig:ht}
\end{figure}
\end{center}
%

Stable period-2 orbits can be observed in Fig.~\ref{fig:ht}(a) for $\beta=0.3$, within the interval 
$0.52 \lesssim \varepsilon \lesssim 0.83$. 
This result is in full agreement with that calculated from the eigenvalues $\lambda(a_k)$ of the matrix 
$\mathbf{F^2 }\mathbf{F^1 }$ (details not shown).
It is apparent that the period-2 family disappears through an inverse period-doubling bifurcation, where a fixed
point is born. 
We already calculated analytically where this bifurcation occurs (see preceding subsection),
namely, $\varepsilon \approx 0.833$, which also agrees with the previous values.

\subsection{Stability of CS period-$p$ trajectories}

In the coupled system, a period-$p$ trajectory (with values $X^1,X^2,\ldots,X^p$) must verify
\begin{eqnarray} 
X^1 &=& (1-\varepsilon) f(X^p) + \varepsilon f( \beta X^{p-1} +[1-\beta] X^p) \\ \nonumber
X^2 &=& (1-\varepsilon) f(X^1) + \varepsilon f( \beta X^{p} +[1-\beta] X^1)  \\  \nonumber
\vdots &=& \vdots \\  \nonumber
X^p &=& (1-\varepsilon) f(X^{p-1}) + \varepsilon f(\beta X^{p-2} +[1-\beta] X^{p-1}) 
\end{eqnarray}
Its stability matrix is given by the product $ \mathbf{F^p } \ldots \mathbf{F^2 }\mathbf{F^1 } $, 
where we have used the natural extension of the notation of previous subsection. 
   
In the particular case $p=3$, a stable family can be observed in Fig.~\ref{fig:ht}(b)
in the interval $0.6077 \lesssim \varepsilon \lesssim 0.6401$ (we checked that this interval coincides with 
the calculation based on the stability matrix).
The latter interval overlaps partially with the family of fixed points 
($0.510 \lesssim  \varepsilon \le 0.625$), indicating the coexistence of two different families of periodic orbits (bistability).

\subsection{Stability of an arbitrary completely synchronized trajectory}

For calculating the stability of long trajectories (periodic, quasiperiodic, chaotic, ...) 
in principle one must resort to purely numerical calculations. 
Consider an arbitrary orbit of length $L$ generated by the map of Eq.~(\ref{2Dmap}),
starting from appropriate initial conditions for selecting the desired orbit. 
$L$ must be large enough for the orbit to fall onto the attractor; 
a transient of length $M$ may be discarded.  
The stability matrix of such an orbit reads  
$\mathbf{F^L }\mathbf{F^{L-1} } \ldots \mathbf{F^{M+2} }\mathbf{F^{M+1} }$.
Upon diagonalization, this matrix product decouples into $N$ matrices of 2$\times$2, 
each one corresponding to one eigenvalue $a_k$ of $\mathbf{A}$.
By analyzing the eigenvalues of all 2$\times$2 matrices we obtain the stability
of the trajectory.

Concerning the possibility of analytical calculations, 
some results can be obtained in the case of chaotic orbits.
In the particular case $\beta=0$ (no delay) each map evolves with the uncoupled local chaotic dynamics. 
All maps are in the same state $X^t$ at time $t$: The full interval $[0,1]$ is a smooth attractor.
The one-step stability matrix reads
\begin{equation}
\mathbf{F}^t = [ (1-\varepsilon) \mathbb{1} + \varepsilon \mathbf{A} ]f'(X^t)  \,. 
\label{matrixF}
\end{equation}
This leads to well-known stability condition (see e.g., \cite{Anteneodo2004,lind04})
\begin{equation}
-1\le {\rm e}^{\lambda_u}[1-\varepsilon(1-a_k)]  \le 1, \;\;\;\mbox{for all $k<N$} \, ,
\label{frontiers}
\end{equation}
where $\lambda_u$ is the Lyapunov exponent of the uncoupled map.
The domain of chaotic synchronization defined by the double inequality (\ref{frontiers}) 
defines the region in the plane $\varepsilon-\alpha$ below the dashed line in Fig.~\ref{fig:PD1}(a) (case $\beta=0$). 
The critical strength $\varepsilon_c$ increases with $\alpha$. 
For instance, for $\alpha=0$ (global coupling, mean field) $\varepsilon_c=0.5$.
The synchronization interval shrinks as $\alpha$ grows and collapses at $\alpha\simeq 0.8$. 
Therefore, too short-range interactions are not able to fully synchronize the system~\cite{Sinha_2000,Anteneodo2004}. 

For $\beta>0$, from a technical point of view, 
the stability matrices acquire the $2 \times 2$ block structure shown in Eq.~(\ref{tildeF}) and it 
is not possible to extract the factor $f'(X^t)$.
At a more intuitive level, as soon as $\beta$ is non-null the bifurcation diagram becomes fractal,
and it is clear that analytical treatments are extremely difficult:
$\beta$ is not a perturbation parameter. \\[5mm]


\section{Two dimensional map}
\label{sec:2Dmap}

\begin{center}
\begin{figure}[b!]
\includegraphics[width=0.98\linewidth,angle=0]{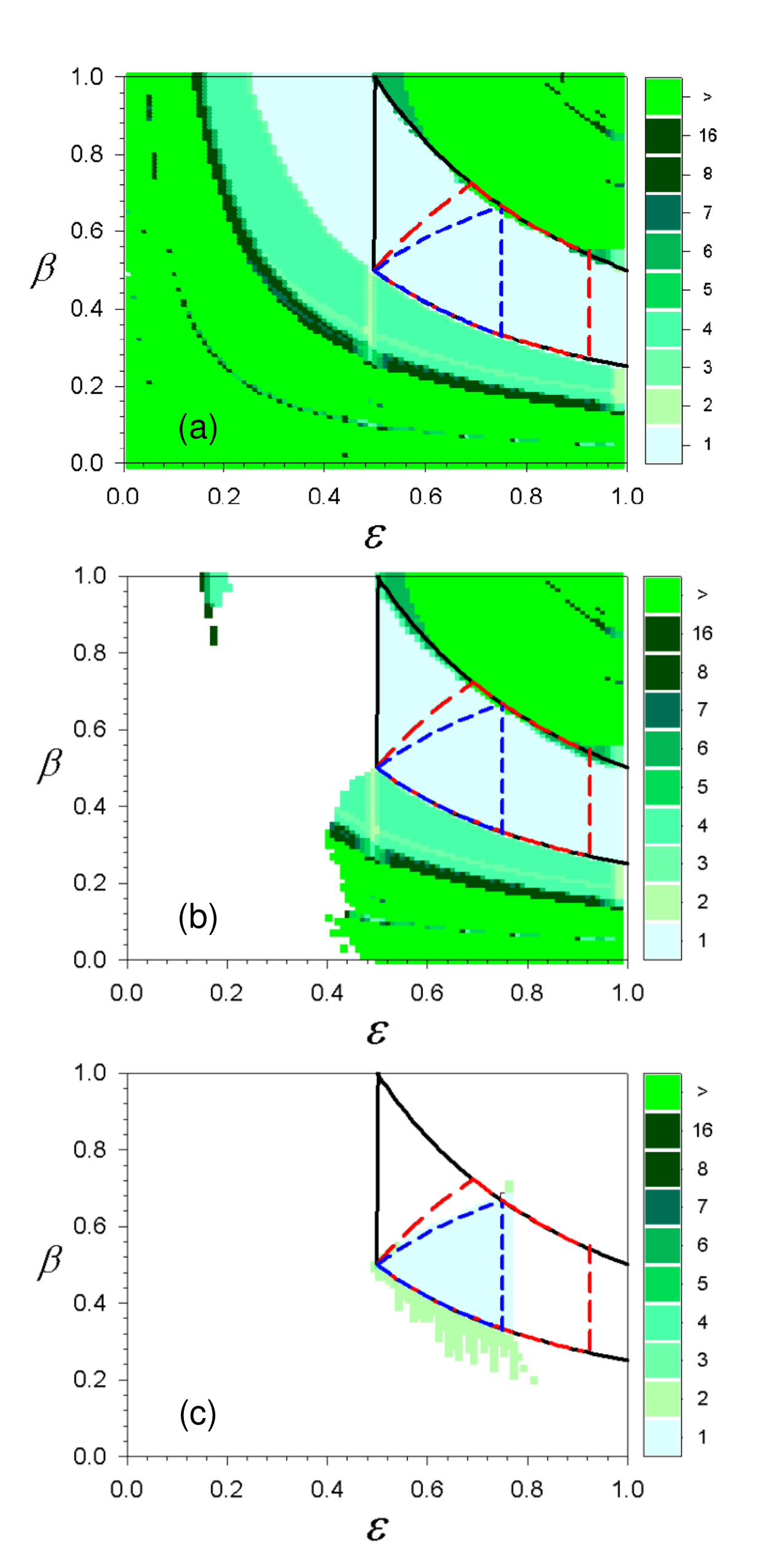} 
\caption{ (Color online)  
(a) Asymptotic states of the map of Eq.~(\ref{2Dmap}).
(b) Stable completely synchronized states of the globally coupled map system ($N=201$, $\alpha=0$).
(c) Same as (b), but for nearest-neighbor couplings ($N=201$, $\alpha=\infty$).
The local dynamics is that of the logistic map $f(x)=4x(1-x)$.
The region of stability of the fixed point $X^0=3/4$ is depicted in light blue.
This region is bounded by the arcs of hyperbola given by $1/4 < \beta \varepsilon < 1/2$
and Eqs.~(\ref{eq:stab1}) [$\alpha=0$ (solid black); $\alpha=2.4$ (long-dashed red); 
$\alpha=\infty$ (short-dashed blue)].   
}
\label{fig:htlog}
\end{figure}
\end{center}
\begin{center}
\begin{figure}[h!]
\includegraphics[width=0.9\linewidth,angle=0]{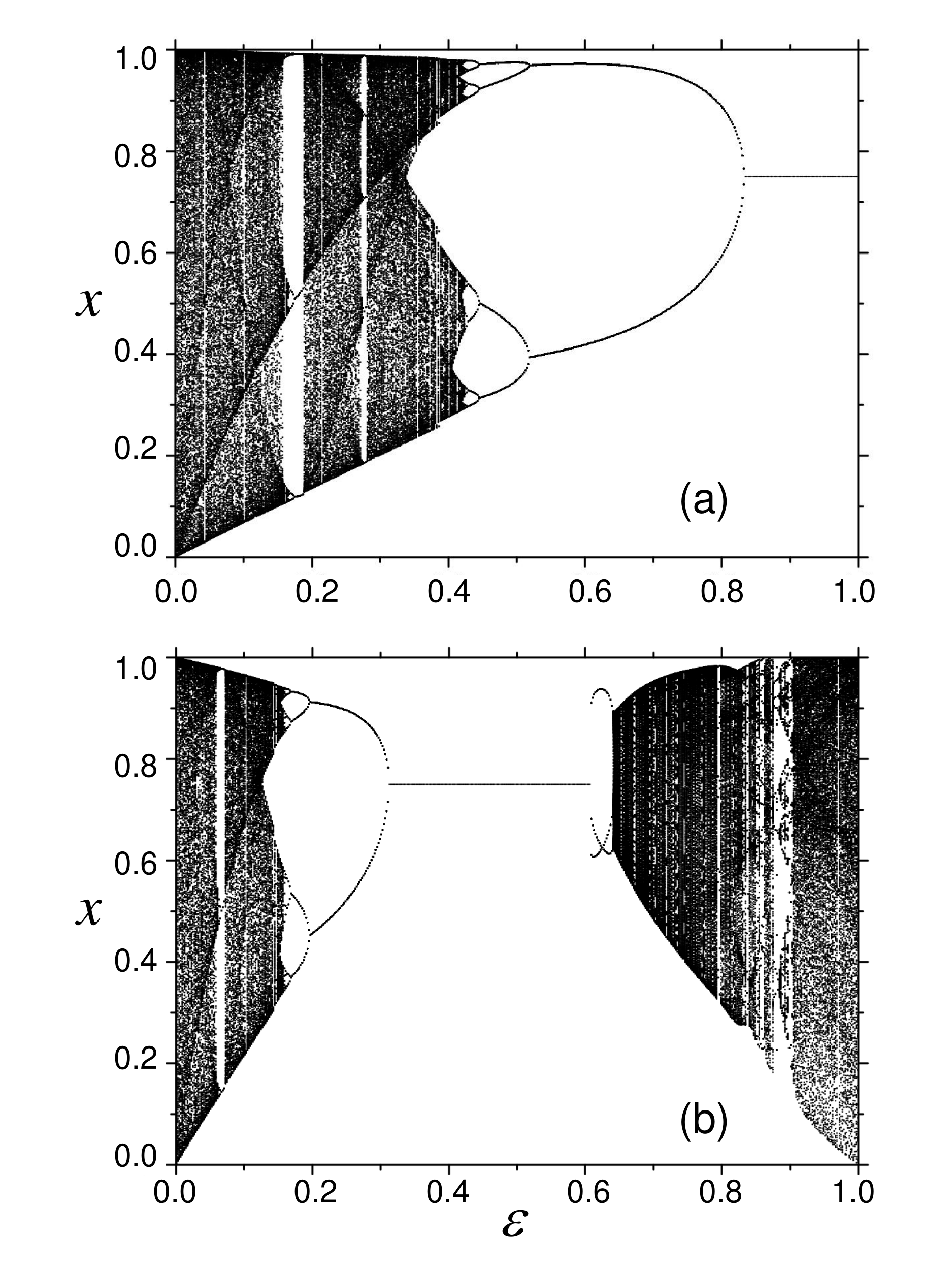}
\caption{ Bifurcation diagram of the completely synchronized map of Eq.~(\ref{2Dmap}) 
for $\beta=0.3$ (a) and $\beta=0.8$ (b).}
\label{fig:bif}
\end{figure}
\end{center}

%
We have shown that  in CSSs  the dynamics is governed by Eq.~(\ref{2Dmap}), 
which is independent on $\alpha$ and $N$, 
although these parameters still rule the stability of CSSs. 
Then, Eq.~(\ref{2Dmap})  which in essence is a 2D map (since it depends on two times) 
determines the structure of bifurcations of CSSs.  
In Fig.~\ref{fig:htlog}(a) we plot the period of the asymptotic solutions 
in the plane $\beta-\varepsilon$ (up to period-16);
longer-period or non-periodic trajectories are represented by green pixels. 
For comparison we have drawn analogous figures for
the system of coupled maps at full coupling, i.e., $\alpha=0$ (b) and
first-neighbor coupling, $\alpha=\infty$ (c).
White regions in Figs.~\ref{fig:htlog}(b)-(c) consist of values of ($\beta,\varepsilon$) 
which do not generate stable CSSs for random initial conditions in [0,1].
Note that white regions spread as $\alpha$ grows. 
In particular, the region of period-1 synchronization 
shrinks as the range of the interaction decreases. 
For $\alpha=\infty$ only a triangle-like stability region remains, corresponding to
low-period orbits (mainly period-one). 
In order to look into the structure of Fig.~\ref{fig:htlog}(a) with greater detail
we plotted bifurcation diagrams at $\beta=0.3$ and $\beta=0.8$ for the 2D-map. 
See Fig.~\ref{fig:bif}(a) and Fig.~\ref{fig:bif}(b), respectively. 
Bifurcation diagrams as a function of $\beta$ for fixed $\alpha$ are qualitatively 
similar to those shown in Fig.~\ref{fig:bif}, since both parameters play a similar role 
for the 2D map. This feature is evident in Fig.~\ref{fig:htlog}(a), 
by observing the approximate symmetry with respect to the diagonal.

A first look at Fig.~\ref{fig:htlog}(b), corresponding to the globally coupled case, shows that most of the synchronized states 
of the system of coupled maps lie above $\varepsilon=0.4-0.5$. 
In fact, a minimum coupling is expected for complete synchronization. 
There is a small island of stability at $\varepsilon \approx 0.2$, and the
existence of other small structures is not discarded. 
However very small regions are difficult to detect numerically because,
presumably, they possess very small basins of attraction, then fine-tuning 
of initial conditions would be required. 
Consider, for instance, the surprising Fig.~\ref{fig:PD1}(f), for $\beta=0.5$.
This figure suggests that only period-1 orbits are stable.
However, the associated bifurcation diagram for the 2D map (not shown) exhibits 
an inverse period doubling cascade for $\varepsilon < 0.5$ (analogous to those
depicted in Fig.~\ref{fig:bif}) and we verified that some of these orbits 
are indeed stable (e.g., a period-2 orbit at $\varepsilon=0.4$ and $\alpha=0.25$). 
So, because of limitations of computing time, Figs.~\ref{fig:PD1} and \ref{fig:htlog} 
only exhibit coarse-grained structures.
Nonetheless, by complementing these figures with fine-detail bifurcation diagrams
we can explain several features of the synchronized states.

%

For weak delays ($\beta < 0.5$), the bifurcation diagrams are logistic-like, i.e.,
typical of one-dimensional maps. 
This can be clearly seen in Fig.~\ref{fig:bif}(a), for $\beta=0.3$. 
However, for stronger delays the dynamics of synchronized states becomes really
two-dimensional. 
As a consequence, e.g., for $\beta=0.8$ (and $\varepsilon > 0.5$), 
we see in Fig.~\ref{fig:bif}(b) bifurcations which are characteristic of 2D maps, 
e.g., Neimark-Sacker bifurcations \cite{thompson}.
Notice that the light-green region observed in Fig.~\ref{fig:PD1}(h) at $\alpha=0.5$ when $\beta=0.8$, 
contains diverse structures, as can be seen in Fig.~\ref{fig:htlog}(a) and also in Fig.~\ref{fig:bif}(b), 
where the additional information provided by the largest Lyapunov exponent indicates that there are 
quase-periodic, long-periodic and also chaotic orbits.

%
\section{Other local dynamics}

In order to test the robustness of our findings, we substituted the logistic map in the local dynamics 
by either cubic or logarithmic maps.

(i) {\em Cubic map}~\cite{cubicmap,lind04c}:
\begin{equation} \label{cubic}
 x \to f(x)=-x^3+ax+b\,,
\end{equation}
with $a=1.5$ and $b=-1$.
Initial conditions were taken random in $[-1,1]$. 
By solving $x=f(x)$, we find the fixed point  $X^0 \approx -1.16537$, with $D^0\approx -2.5742$. 

(ii) {\em Logarithmic map}~\cite{cosenza}:
\begin{equation} \label{logarithmic}
 x \to f(x)=c+ \ln|x| \,,
\end{equation}
with $c=0.0$. 
In this case the initial conditions for each map were random in $[-10,10]$. 
This map has the fixed point $X^0 \simeq -0.567143$, 
with $D^0\equiv f'(X^0)= 1/X^0\simeq  -1.763223$. 
The longitudinal stability condition is  $0.216 \lesssim \beta\varepsilon \lesssim 0.567$.
%

\begin{center}
\begin{figure}[h!]
\includegraphics[width=0.9\linewidth,angle=0]{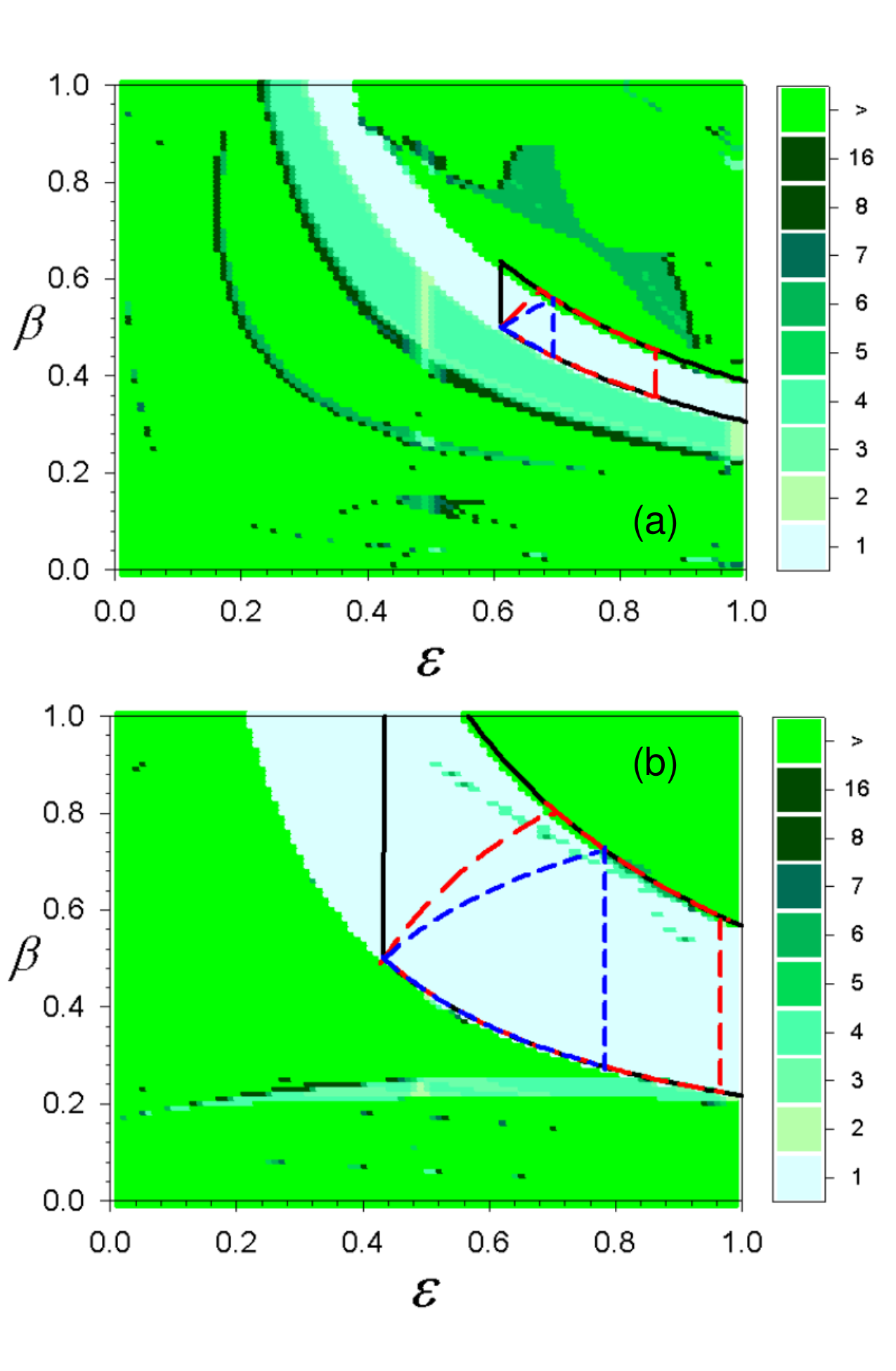}
\caption{ (Color online)  
Phase diagram in the parameter plane $(\varepsilon,\alpha)$ 
for the cubic (a) and logarithmic (b) delayed 2D maps 
[cf. Fig.~\ref{fig:htlog}(a)].
The color scale shows the period of the orbits in the 2D maps, 
starting from random initial values in $[-1,1]$ and $[-10,10]$ 
for the cubic and logarithmic maps, respectively. 
The lines, given by Eqs.~(\ref{frontiers1},\ref{frontiers2}), 
delimit the region of period-1 complete synchronization.
}
\label{fig:other2D}
\end{figure}
\end{center}

The cubic map is not unimodal and 
the logarithmic map is neither unimodal nor bounded~\cite{cosenza}.
The  cubic map has bifurcation diagrams similar to the 
logistic one, both as a function of $a$ and $b$. 
In the logarithmic map bifurcation diagram, as a function of $c$, 
there is a single fixed point that loses stability for $c\in[-1,1]$ and   
chaos emerges,  without windows of regularity.

We built phase diagrams for these maps, shown in Fig.~\ref{fig:other2D}, 
where the frontiers for the fixed point stability in CS states are highlighted. 

The longitudinal stability condition, that generalizes Eq.~(\ref{cond1}) 
for the logistic map, is 
\begin{equation}
\frac{1+D^0}{2D^0}<   \beta\varepsilon  < -\frac{1}{D^0} .
\label{frontiers1}
\end{equation}
These inequalities define the band of CS period-1 orbits observed in Fig.~\ref{fig:other2D}.

The transversal condition generalizing Eq.~(\ref{eq:stab1}) is given by
\begin{equation}
\frac{1+D^0}{2D^0[ 1-a_{\rm min} (1 -2 \beta  ) ]} < \varepsilon <  \frac{D^0-1}{D^0 ( 1 - a_{\rm min}  )} \, . 
\label{frontiers2}
\end{equation}
This condition depends on the range of the interactions, which restricts the band of period-1 orbits 
as the range decreases (see Figs.~\ref{fig:htlog} and \ref{fig:other2D}).

\begin{center}
\begin{figure}[h!]
\includegraphics[width=0.8\linewidth,angle=0]{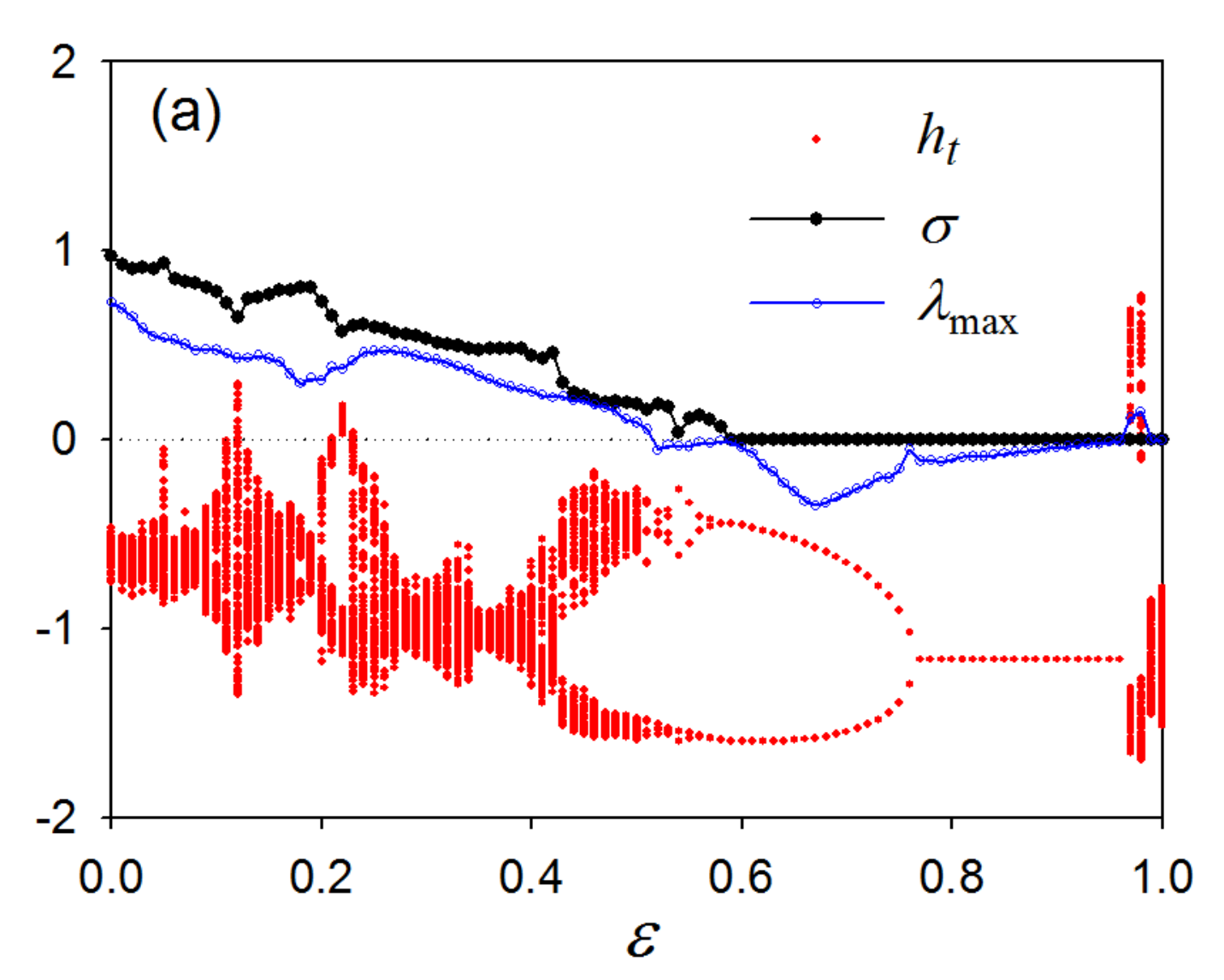} 
\includegraphics[width=0.8\linewidth,angle=0]{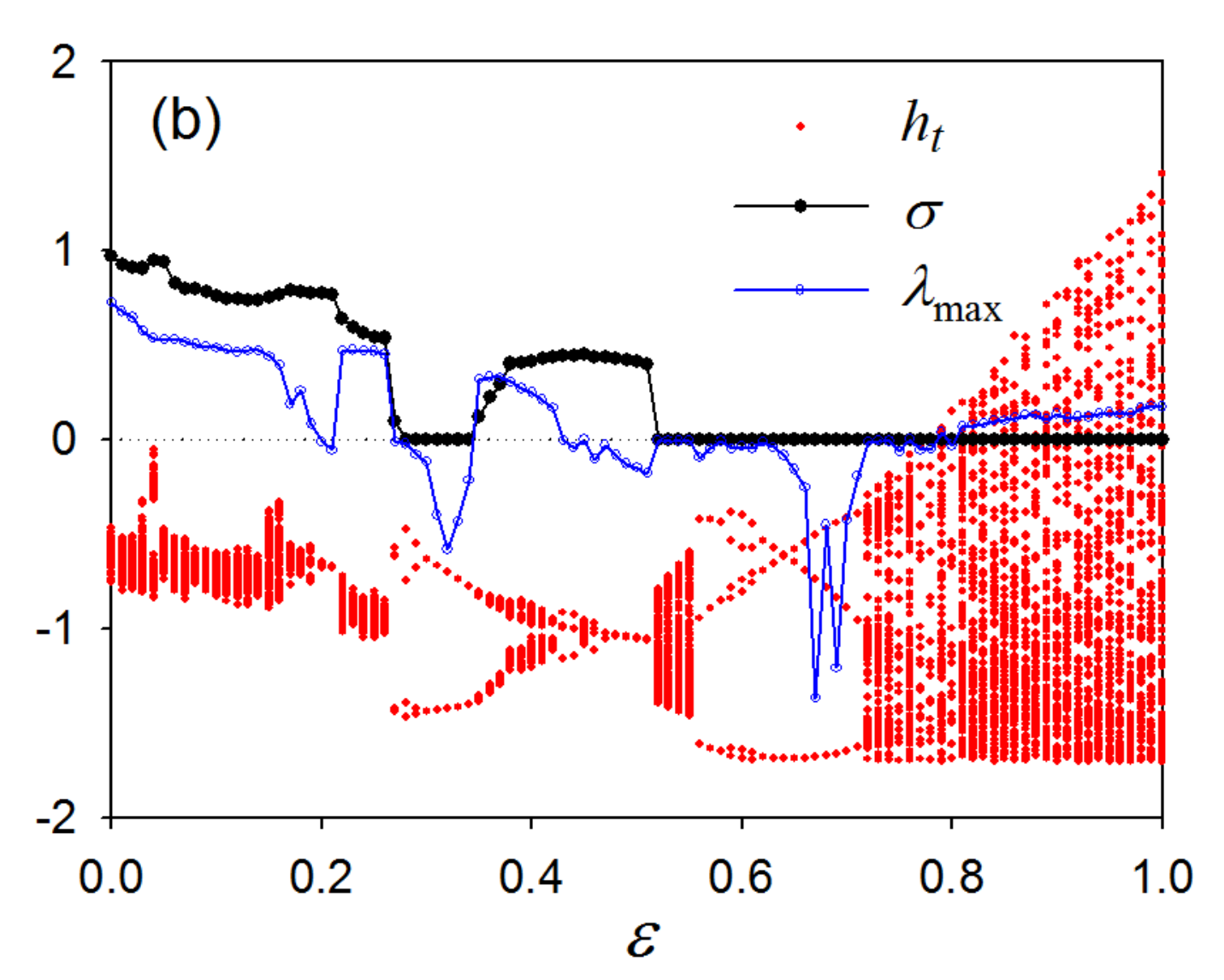}  
\caption{ (Color online)  
Cubic maps. Order parameter $\sigma$ (black dots) and  mean field $h_t$ (red light dots) vs. coupling $\varepsilon$, 
for $\beta=$ 0.4 (a) and $\beta=$ 0.8 (b). 
In all cases the array size is  $N=201$ and $\alpha=0.5$. 
}
\label{fig:htcubic}
\end{figure}
\end{center}

\begin{center}
\begin{figure}[h!]
\includegraphics[width=0.8\linewidth,angle=0]{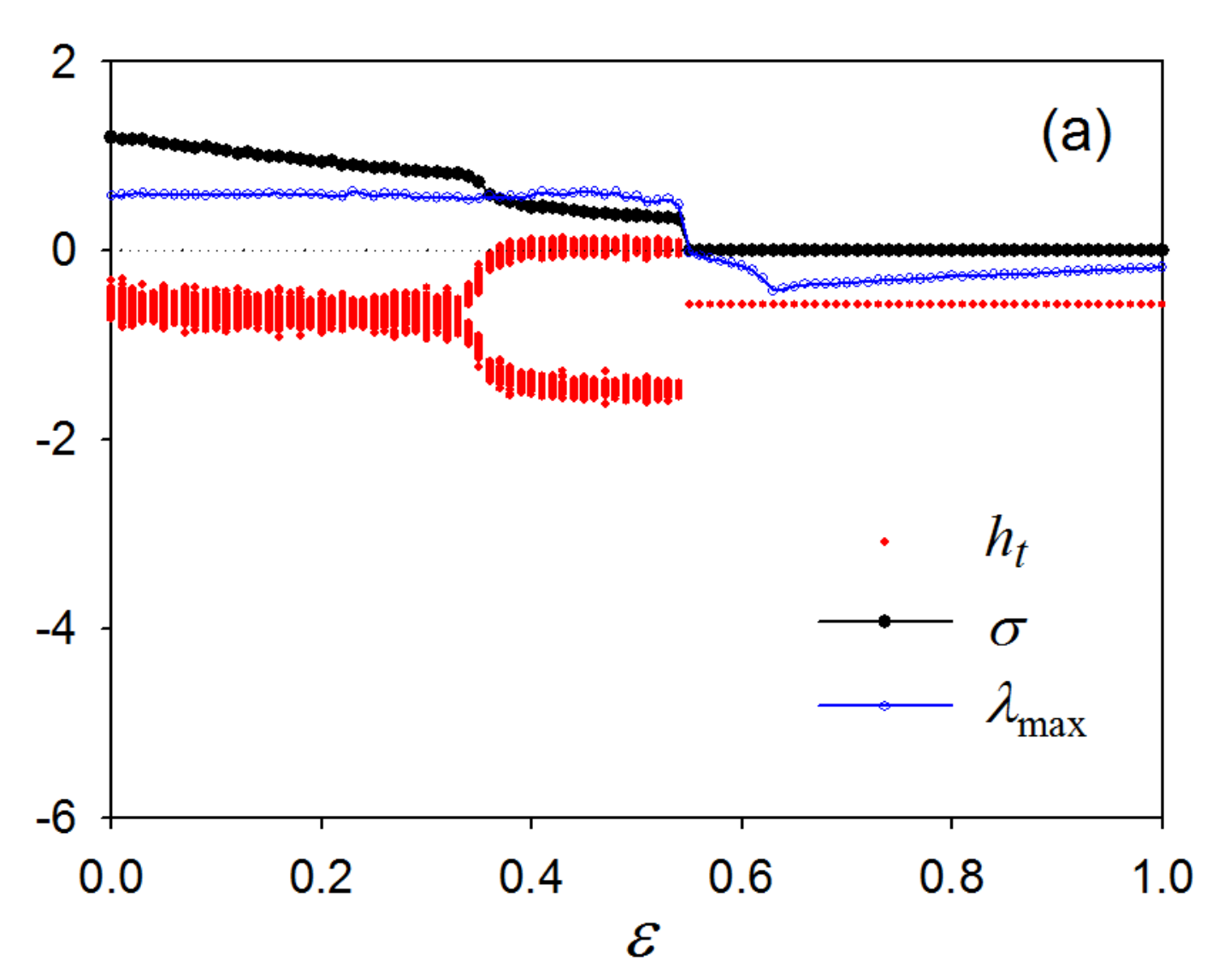} 
\includegraphics[width=0.8\linewidth,angle=0]{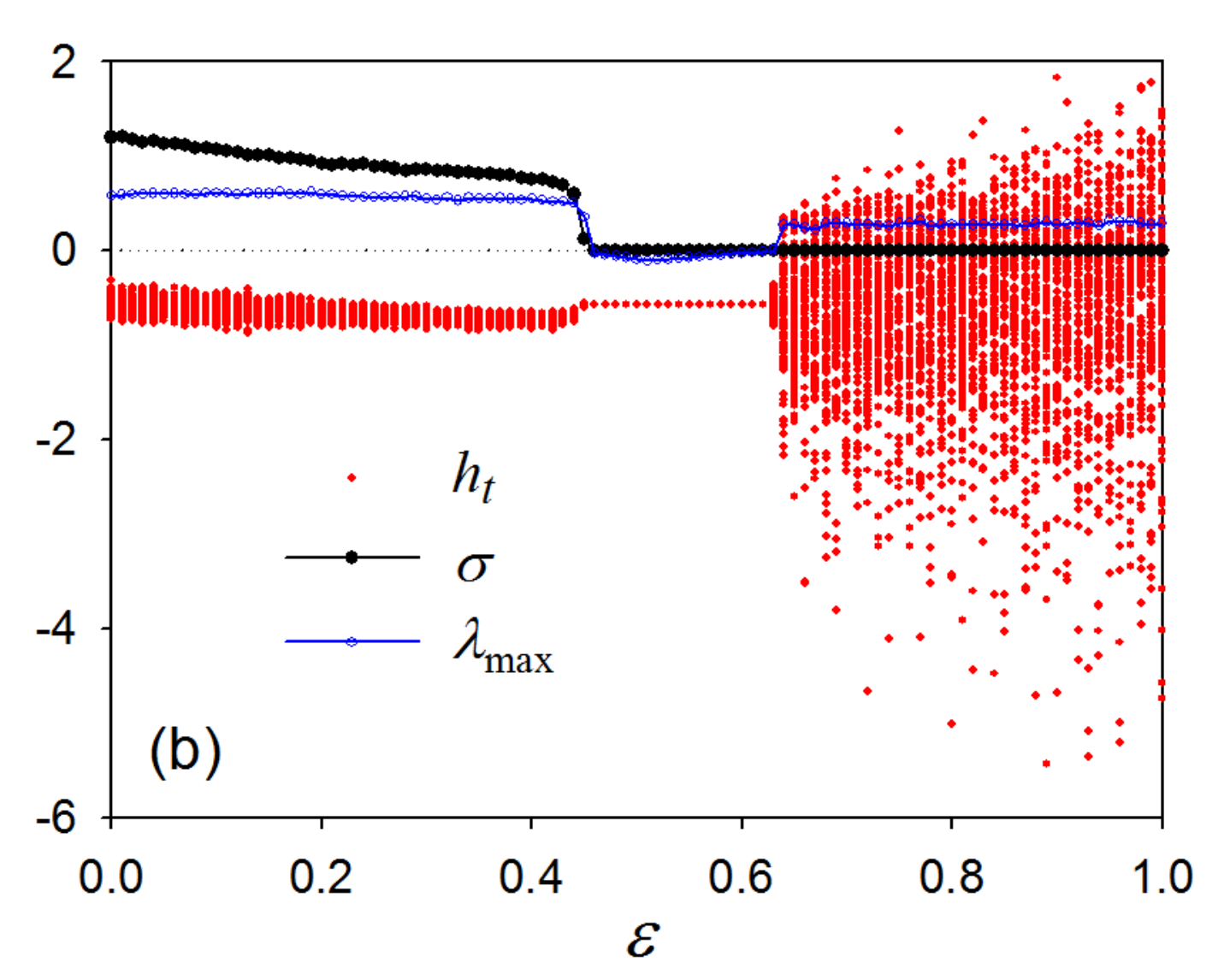}  
\caption{ (Color online)  
Logarithmic maps. Order parameter $\sigma$ (black dots) and  mean field $h_t$ (red light dots) vs. coupling $\varepsilon$, 
for $\beta=$ 0.4 (a) and $\beta=$ 0.9 (b). 
In all cases the array size is  $N=201$ and $\alpha=0.5$. 
}
\label{fig:htlogarit}
\end{figure}
\end{center}

The effect of the delay in coupled cubic maps is qualitatively similar to that found for the logistic maps. 
In fact the phase diagram of the cubic map, depicted in Fig.~\ref{fig:other2D}(a), 
is a deformation of Fig.~\ref{fig:htlog}(a). 
In both cases the band of CS period-1 orbits separates two regions 
with either one or two dimensional behavior, as one can see in the bifurcation diagrams 
in Fig.~\ref{fig:htcubic}. 
Incidentally, the structures which appear in Fig.~\ref{fig:other2D}(a) 
remind us the shrimp-like structures found in the parameter 
space of the H\'enon map~\cite{gallas93,beims11}. 
However, in order to certify the existence of such structures, we should select very 
small regions with a much finer discretization, a study that although interesting is beyond 
our present scope.

In the logarithmic case [Fig.~\ref{fig:other2D}(b)] there is also a band of period-1 orbits, 
and in the region of small $\beta\varepsilon$, 
complete synchronization of periodic windows with period larger than 1 appear (unstable in the uncoupled map), 
like for the logistic and cubic dynamics. 
The thin colored region within the period-1 band contains orbits of higher period. 
For instance, we checked that for $\beta=0.8$ and $\varepsilon \approx 0.65$ 
there are period-3 orbits coexisting with period-1 ones 
---a case of bistability also observed in the logistic system.
However, for large $\beta\varepsilon$, the  behavior typical of bidimensional maps, with structures such as 
limit cycles, does not emerge. Bifurcation diagrams are illustrated in Fig.~\ref{fig:htlogarit}, 
for two different values of $\beta$.
%


\section{Final remarks}
 
We selected a coupled system that scans continuously from synchronized to one-time delayed dynamics.
We used the logistic map local dynamics as paradigm, but general analytical expressions were obtained 
and other maps were also simulated. 

We exhibited the phenomenology that appears in the route of increasing contribution 
of the delays (increasing $\beta$), focusing on complete synchronization. 
Basically, we distinguish two scenarios. 
One where delays are not dominant 
(small enough $\beta$, namely $\beta<0.5$ in the case of the logistic map), 
in which case the bifurcation diagram of the coupled maps 
is qualitatively similar to that for the local dynamics, although deformed. 
That is, for small contribution of the delays, 
the dynamics of completely synchronized states remains essentially one-dimensional. 
The other scenario appears when the dynamics is delay-dominated 
(large enough $\beta$, i.e., $\beta>0.5$ for the logistic map), 
where a scenario typical of 2D maps emerges.
Noticeably, the transition between both kind of behavior occurs through a fixed point synchronization.

Although in many cases delays regularize the dynamics, we clearly see that they can also   
have the opposite effect, depending on the coupling parameters. 
That is, for small contribution of delays, by increasing $\beta$, 
chaos (together with the full logistic-like bifurcation diagram)
shrinks towards $\varepsilon=0$ (see Fig.~\ref{fig:bif}).
But for large contribution of delays, when increasing $\beta$, chaos can originate 
from the breakdown of limit cycles within the 2D scenario. 
The same portrait was observed for the cubic map. 

Despite 2D structures not being observed in the case of local logarithmic dynamics, 
also here chaos is broken for weak delays and created by strong ones.

We have also shown the interplay of the delays and the range of the interactions. 
Long-range generically aids complete synchronization, which will not be possible for short-range, 
in non-delayed dynamics. 
However, moderate contribution of delays allows complete synchronization even for nearest neighbors, 
as displayed, for instance, in Fig.~\ref{fig:htlog}.

{\bf Acknowledgments:} 
We acknowledge Brazilian agencies CNPq and FAPERJ for financial support.


\end{document}